\documentclass[aps,10pt,prd,notitlepage,showpacs,nofootinbib,superscriptaddress]{revtex4-1}

\usepackage{graphicx}
\usepackage[utf8]{inputenc} 
\usepackage{amsmath}
\usepackage{amssymb}
\usepackage{float}
\usepackage{comment}
\usepackage{slashed}
\usepackage[normalem]{ulem}
\usepackage{cancel}
\makeatother
\usepackage{multirow}
\usepackage{rotating}
\usepackage[usenames,dvipsnames]{color}
\usepackage[colorinlistoftodos]{todonotes}
\usepackage[colorlinks=true,citecolor=blue,urlcolor=blue, pdfborder={0 0 0}]{hyperref}
\usepackage[normalem]{ulem}
\definecolor{darkred}{rgb}{0.6,0,0}
\usepackage[colorinlistoftodos]{todonotes}
\definecolor{linkcolor}{rgb}{0,0,0.5}

\usepackage{soul}
\definecolor{mightnightblue}{RGB}{25,25,112}

\definecolor{brown}{rgb}{0.59, 0.29, 0.0}

\begin{document}

\title{\color{BrickRed}Laboratory limits on the annihilation or decay of dark matter particles}
\author{Teresa Marrodán Undagoitia}
\email{teresa.marrodan@mpi-hd.mpg.de}
\affiliation{Max-Planck-Institut f\"ur Kernphysik, Postfach 103980, D-69029
Heidelberg, Germany}
\author{Werner Rodejohann}
\email {werner.rodejohann@mpi-hd.mpg.de}
\affiliation{Max-Planck-Institut f\"ur Kernphysik, Postfach 103980, D-69029
Heidelberg, Germany}
\author{Tim Wolf}
\email {tim.wolf@mpi-hd.mpg.de}
\affiliation{Max-Planck-Institut f\"ur Kernphysik, Postfach 103980, D-69029
Heidelberg, Germany}
\author{Carlos E.\ Yaguna}
\email{carlos.yaguna@uptc.edu.co}
\affiliation{Escuela de Física, Universidad Pedagógica y Tecnológica de Colombia, Avenida Central del
Norte 39-115, Tunja, Colombia}

\begin{abstract}\noindent
Constraints on the indirect detection of dark matter are usually obtained from observations of astrophysical objects -- the Galactic Center, dwarf galaxies, M31, etc. Here we propose instead to  
look for the annihilation or decay of dark matter particles taking place inside detectors searching \emph{directly} for dark matter or in large neutrino experiments. We show that the data from XENON1T and Borexino set limits on the annihilation and decay rate of dark matter particles with masses in the keV to few MeV range. All  relevant final states are considered: annihilation into $\gamma\gamma$ and $e^-e^+$, and decays into $\gamma\gamma$, $\gamma\nu$ and $e^-e^+$. The expected sensitivities in XENONnT, DARWIN, JUNO and THEIA are also computed. Though weaker than current astrophysical bounds, the laboratory limits (and projections) obtained are free from the usual astrophysical uncertainties associated with $J$-factors and unknown backgrounds, and may thus offer a complementary probe of the dark matter properties. We point out that current and future (astro)particle physics detectors  might also be used to set analogous limits for different decays and dark matter masses above a few MeV.  
\end{abstract}
\date{\today }
\maketitle
\section{Introduction}
\noindent 
Constraints on dark matter (DM) properties can be obtained in different ways. The main methods are direct detection\,\cite{Undagoitia:2015gya,Billard:2021uyg}, in which the cross section of DM particles with nuclei or electrons is measured, and indirect detection\,\cite{Klasen:2015uma}, in which the decay or annihilation rate of DM particles into Standard Model (SM) particles is measured. An additional method is the production of DM particles at colliders. In that case the signatures are events with missing energy, which is attributed to DM particles being produced and escaping the detector\,\cite{Kahlhoefer:2017dnp}. 
\vskip 0.2cm

Direct detection is subject to many sophisticated experiments, using different target nuclei and measurement techniques. Looking for tiny recoils stemming from a DM-nucleus or DM-electron scattering, the constrained parameters are the mass of the DM particle and its cross section with the target particle. Indirect detection is typically performed in a parasitic way in telescopes aiming to perform 
astrophysical measurements. In this case, the DM annihilation or decay rate in any SM particle is constrained as a function of the DM mass, from looking directly for those SM particles, photons being radiated off them, or their decay products. Inherent to such measurements are astrophysical uncertainties such as the DM density along the line of sight (called $J$-factor), or SM processes in the observed complex objects that generate very similar signatures. 
Objects that are particularly popular in what regards observing them are the galactic center, dwarf galaxies, or the Andromeda galaxy.
\vskip 0.2cm

In this paper, we suggest to search for events where the DM decay or annihilation takes place inside Earth-based detectors. The requirements for such a detector are: the ability to reconstruct the energy deposited by particles (spectroscopy), a large target volume and ideally a very low background. Our study is focused on two-phase liquid xenon experiments and organic scintillation detectors. These technologies provide extremely clean environments, as demonstrated by world-leading direct detection results and background rates by XENON1T\,\cite{Aprile:2018dbl} and by the measurements of all solar neutrino components in Borexino\,\cite{Agostini:2017ixy}.
The energy depositions expected in these detectors from DM interactions or solar neutrinos are in the keV or MeV regime, and the experiments are optimized to look for such signals. 
 Therefore, our constraints on DM annihilation and decay will be in this mass range. While it is expected that such laboratory limits on indirect detection rates will be much weaker than the ones from astrophysical observations of whole galaxies, the approach proposed here has several advantages: $i)$ the experimental conditions are under control so the backgrounds are known; $ii)$  the signal does not suffer from astrophysical uncertainties such as the $J$-factor; $iii)$ significant improvements in the limits can be foreseen from the upcoming generation of experiments. 
\vskip 0.2cm

We consider annihilation processes into $e^- e^+$ and  $\gamma\gamma$, and  decays into $e^- e^+$, $\gamma\gamma$ and $\gamma\nu$. We shall analyze public XENON1T data\,\cite{Aprile:2020tmw, Aprile:2020yad}
and extrapolate the limits to upcoming XENONnT\,\cite{Aprile:2020vtw} and DARWIN\,\cite{Aalbers:2016jon} data, making different assumptions on the background rates in those experiments. The sensitivity of the upcoming LZ experiment\,\cite{Akerib:2015cja} should be very similar to that of XENONnT.
We employ as well the public data from Borexino\,\cite{Agostini:2018uly} to derive results, and make projections for JUNO\,\cite{Djurcic:2015vqa} and THEIA\,\cite{Askins:2019oqj} using their predicted background spectra.

\vskip 0.2cm

This paper is built up as follows: first, in Sec.\ \ref{sec:signal} we discuss the annihilation and decay processes. In Sec.\,\ref{sec:Response and BG} we describe general features of signal production and the background spectra used for the analysis.
In Sec.\ \ref{sec:proc_results} we derive limits on existing and expected future data for the signatures under study, namely DM annihilation and DM decay. We describe these physics processes in detail and explain the  procedure of limit setting. A discussion of our results is presented in Sec.\ \ref{sec:dis}.

\section{\label{sec:signal}Annihilation and decay processes}
\noindent
We consider DM annihilation and decay processes  that may take place inside  terrestrial detectors. Their respective rates (number of events per unit volume and unit time) are given by

\begin{align}
\label{eq:anni}
    \mathcal{R}&=\frac{\langle\sigma v\rangle}{2}\, n_{\rm DM}^2=\frac{\langle\sigma v\rangle}{2m_\chi^2}\, \rho_{0,\rm DM}^2\quad \mathrm{(Annihilation)},\\
    \mathcal{R}&=\Gamma_{\rm DM}\, n_{\rm DM}=\frac{\Gamma_{\rm DM}}{m_\chi}\, \rho_{0,\rm DM}\quad \mathrm{(Decay)},\label{eq:decay}
\end{align}
where $\langle\sigma v\rangle$ and $\Gamma_{\rm DM}$ are respectively the annihilation and decay rates into a given final state, $m_\chi$ is the mass of the DM particle, and $\rho_{0,\rm DM}$ is the local DM density, which we take to be $0.3$\,GeV/cm$^3$\,\cite{Green:2017qoa}. Our goal is then to set limits on $\langle\sigma v\rangle$ and $\Gamma_{\rm DM}$ as a function of $m_\chi$ for all possible final states. Obviously, the smaller the DM mass, the larger is the rate and the stronger the constraint. 
Given the relatively small volume of typical  detectors, meaningful bounds can be derived only for light DM particles (keV to MeV). 
Notice that, besides the particle physics quantities we are interested in,  the event rates  depend only on the \emph{local} DM density. Therefore, they are insensitive to the usual large uncertainties associated with the distribution of DM inside astrophysical objects, typically encoded in the so-called $J$-factors\,\cite{Strigari:2013iaa}. Moreover, unlike direct detection rates, equations (\ref{eq:anni}) and (\ref{eq:decay}) do not depend on the DM velocity distribution.
\vskip 0.2cm

The experiments considered in this work are sensitive to energy depositions between a few keV and a few MeV. 
For DM annihilations and decays, the total energy deposition is of the same order as $m_\chi$, which implies that the range of DM masses where bounds can be derived is $\mathrm{keV}\lesssim m_\chi\lesssim \mathrm{MeV}$.  
\vskip 0.2cm

The possible final states from annihilation or decay are quite limited for DM masses below a few MeV, as the only accessible SM particles  are photons, neutrinos, and electrons/positrons. 
In this analysis, we consider two DM annihilation reactions (into $\gamma\gamma$ and $e^-e^+$) and three decay modes ($\gamma\gamma$, $\gamma\nu$, and $e^-e^+$). We remain agnostic about the nature of the DM particle, but note that decays into a $\gamma\nu$ ($\gamma\gamma$) final state resemble a keV-scale sterile neutrino \cite{Abazajian:2017tcc,Boyarsky:2018tvu}, axion \cite{Chadha-Day:2021szb} or axion-like particle  \cite{Irastorza:2018dyq} signals. Our results apply to many of the countless DM candidates proposed so far. 
\vskip 0.2cm

\section{\label{sec:Response and BG}Detector response and background spectra}
\noindent
Charged particles deposit their energy in liquid xenon producing thereby scintillation and ionization. In dual-phase time-projection chambers, free electrons can be collected and amplified via proportional scintillation in the gas phase on top of the liquid target\,\cite{Aprile:2017aty}. Therefore, both scintillation photons and the ionization signals are read out using photo-sensors. These two signals can be combined\,\cite{Aprile:2020yad} to improve the energy resolution of the detector. In organic liquid scintillators, charged particles deposit their energy similarly exciting the medium. In the de-excitation process, emitted photons propagate towards the walls of the detector where photo-sensors are located. While in liquid xenon detectors, the charge signal can be used to identify multiple scatterings, in organic scintillators this is not possible. Interactions by mono-energetic $\gamma$-particles appear in the latter as mono-energetic peaks resulting from their full absorption.
In this work, we consider annihilation and decay reactions in which observable electrons, positrons and $\gamma$-rays are created. The $\gamma$-rays have to interact in the sensitive volume in order to leave a measurable signal. For the energy range considered in this work, they  interact either via photo-absorption or via Compton scattering. 
\vskip 0.3cm

\noindent
{\bf Energy resolution:}\\
The finite energy resolution of the detectors is taken into account when modeling the signals.
We parametrize the energy resolution of the corresponding signals via the equation
\begin{equation}\label{eq:E_res}
\frac{\sigma(E\,[\textrm{keV}])}{E} = \frac{a}{\sqrt{E\,[\textrm{keV}]}} + b.
\end{equation}
\vskip 0.2cm
In XENON1T, the parameters are determined to be  $a = 31.3$ and $b = 0.17$\,\cite{Aprile:2020yad}.
As our annihilation and decay signatures involve $\gamma$-rays, we use for those a slightly different parametrization which takes into account the fact that above $\sim 350$\,keV $\gamma$-rays undergo multiple scatters. To obtain this slightly different parametrization (with the parameters $a = 31$ and $b = 0.19$), 
we fit the single scatter points below 350\,keV  and the multiple scatters above that energy (data from figure 6 in Ref.\,\cite{Aprile:2020yad}).
\vskip 0.2cm

For the organic liquid-scintillation detectors considered here (Borexino, JUNO and THEIA), we parametrize the energy resolution as in equation\,(\ref{eq:E_res}) setting $b=0$. For the Borexino experiment, we fit the mono-energetic peaks produced by $\gamma$ calibration sources as shown in figure 17 of~\cite{Back:2012awa} with Gaussian functions. We extract the corresponding resolution $\sigma (E)$ and obtain a fit value for equation\,(\ref{eq:E_res}) of $a=169$. For the upcoming/future detectors, we take the expected resolution values as found in the literature corresponding to $3\%/\sqrt{E\,[\textrm{MeV}]}$ and $7\%/\sqrt{E\,[\textrm{MeV}]}$ 
for JUNO\,\cite{Djurcic:2015vqa} and THEIA\,\cite{Askins:2019oqj}, respectively. Table\,\ref{tab:summ} summarizes the energy resolution for all experiments considered here.
\vskip 0.3cm

\noindent
{\bf Efficiency and exposure:}\\
While for organic scintillators we assume a detector efficiency of 100\%, for the xenon-based detectors we take an energy-dependent efficiency to be consistent with the published data (figure\,2 of \cite{Aprile:2020tmw}). The latter has an efficiency loss towards low energies (below $\sim 4$\,keV) and at high energies is flat at the level of 90\%.
\vskip 0.2cm

Our analysis derives results based on the 0.65\,tonne$\cdot$y data of the XENON1T experiment\,\cite{Aprile:2020tmw}, and gives expected sensitivities for its successor XENONnT\,\cite{Aprile:2020vtw} and the future DARWIN experiment\,\cite{Aalbers:2016jon}. The signals in these detectors scale with the exposure, the product of the fiducial mass (mass used for data analysis) and the measuring time. The considered exposures are 20\,tonnes$\cdot$y and 200\,tonnes$\cdot$y for XENONnT and DARWIN, respectively. Note however that for DM annihilation or decay inside these detectors, the volume (and not the mass) is relevant, and therefore we divide the exposure by the xenon density ($\rho = 2.85$\,g/cm$^3$).
To derive results for Borexino, we take an exposure of 1291\,days and 71.3\,tonnes corresponding to a total of 252\,tonne$\cdot$year\,\cite{Agostini:2018uly}. Note that for deposited energies above $\sim 3$\,MeV energies, the fiducial volume of these detectors can be enlarged allowing for significantly larger exposures\,\cite{Agostini:2017cav}. This is however not considered in our study. For JUNO and THEIA, we assume 100\,ktonne$\,\cdot\,$year and 300\,ktonne$\,\cdot\,$year, respectively, in accordance with the expected exposures mentioned in the corresponding publications\,\cite{Djurcic:2015vqa,Askins:2019oqj}. 
 While the density for THEIA is assumed to be 1\,g/cm$^3$, we take for Borexino 0.89\,g/cm$^3$\,\cite{Borexino:2019gps} and for JUNO 0.86\,g/cm$^3$, as reported in\,\cite{Djurcic:2015vqa}.
Table\,\ref{tab:summ} summarizes the parameters employed for each experiment.

\begin{table*}[h]
\caption{Summary of parameters for the experiments considered in this study. Note that for this analysis the exposure given here has to be divided by the detector material's density. 
 }\label{tab:summ}
\begin{center}
\begin{tabular}{c|c|c|c}
	\hline
Experiment ~ & ~ Energy resolution (at 1\,MeV) ~ & ~ Exposure & Density \\
		\hline
XENON1T & 1.2\%\,  &  0.65\,tonne\,$\cdot$\,y  & 2.85\,g\,$\cdot\,\text{cm}^{-3}$ \\
XENONnT & 1.2\%\,  &  20\,tonne\,$\cdot$\,y &  2.85\,g\,$\cdot\,\text{cm}^{-3}$ \\
DARWIN & 1.2\%\,  &  200\,tonne\,$\cdot$\,y &  2.85\,g\,$\cdot\,\text{cm}^{-3}$ \\
Borexino & 5.3\%  &  252\,tonne\,$\cdot$\,y  & 0.89\,g\,$\cdot\,\text{cm}^{-3}$ \\
JUNO & 3\%  &  100\,ktonne\,$\cdot$\,y &  0.86\,g\,$\cdot\,\text{cm}^{-3}$\\
THEIA & 7\%  &  300\,ktonne\,$\cdot$\,y &  1.0\,g\,$\cdot\,\text{cm}^{-3}$ \\
	\hline
\end{tabular}
\end{center}
\end{table*}
\vskip 0.3cm

\begin{figure}[t]
    \centering
    \includegraphics[width=0.48\textwidth]{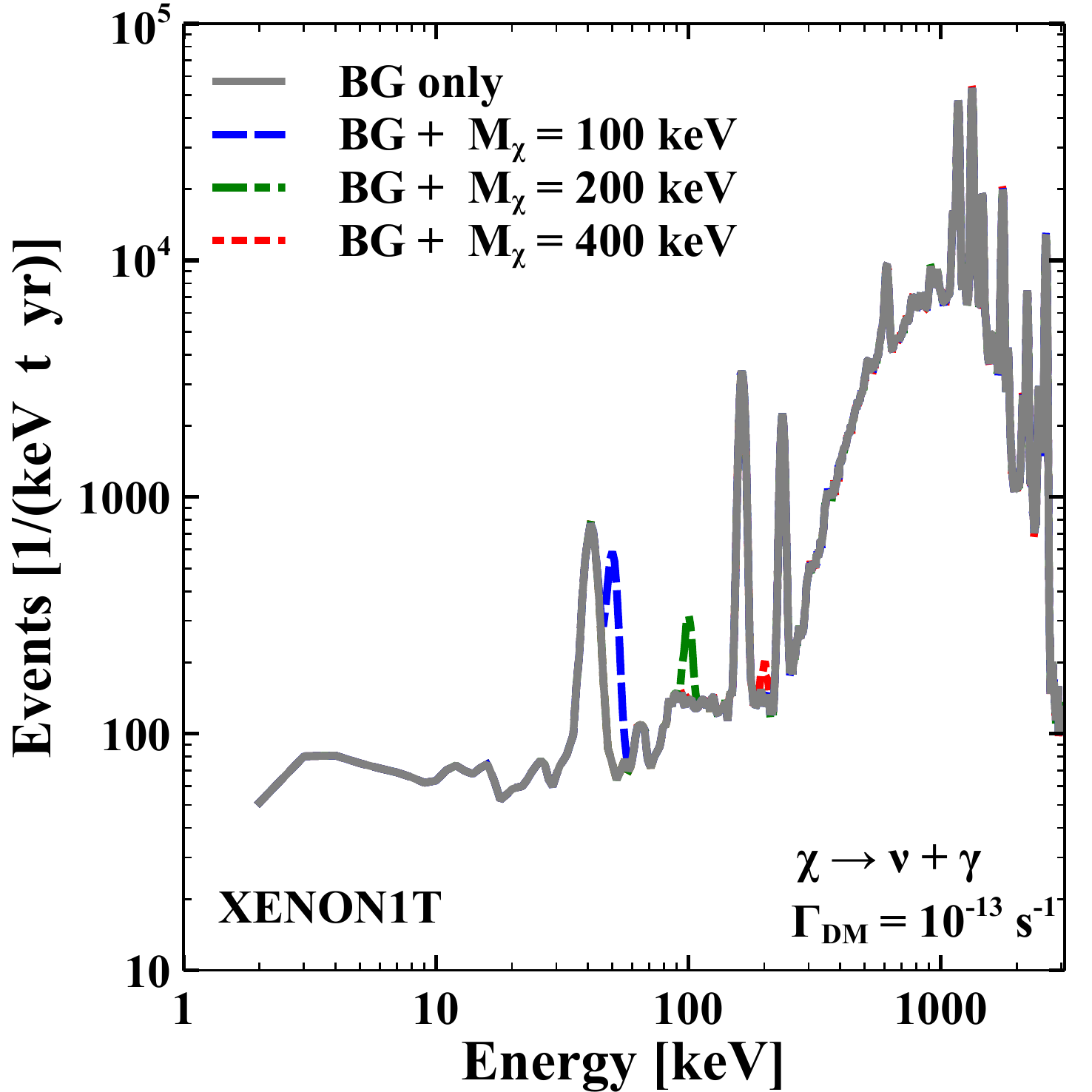} \hspace{3mm}
    \includegraphics[width=0.48\textwidth]{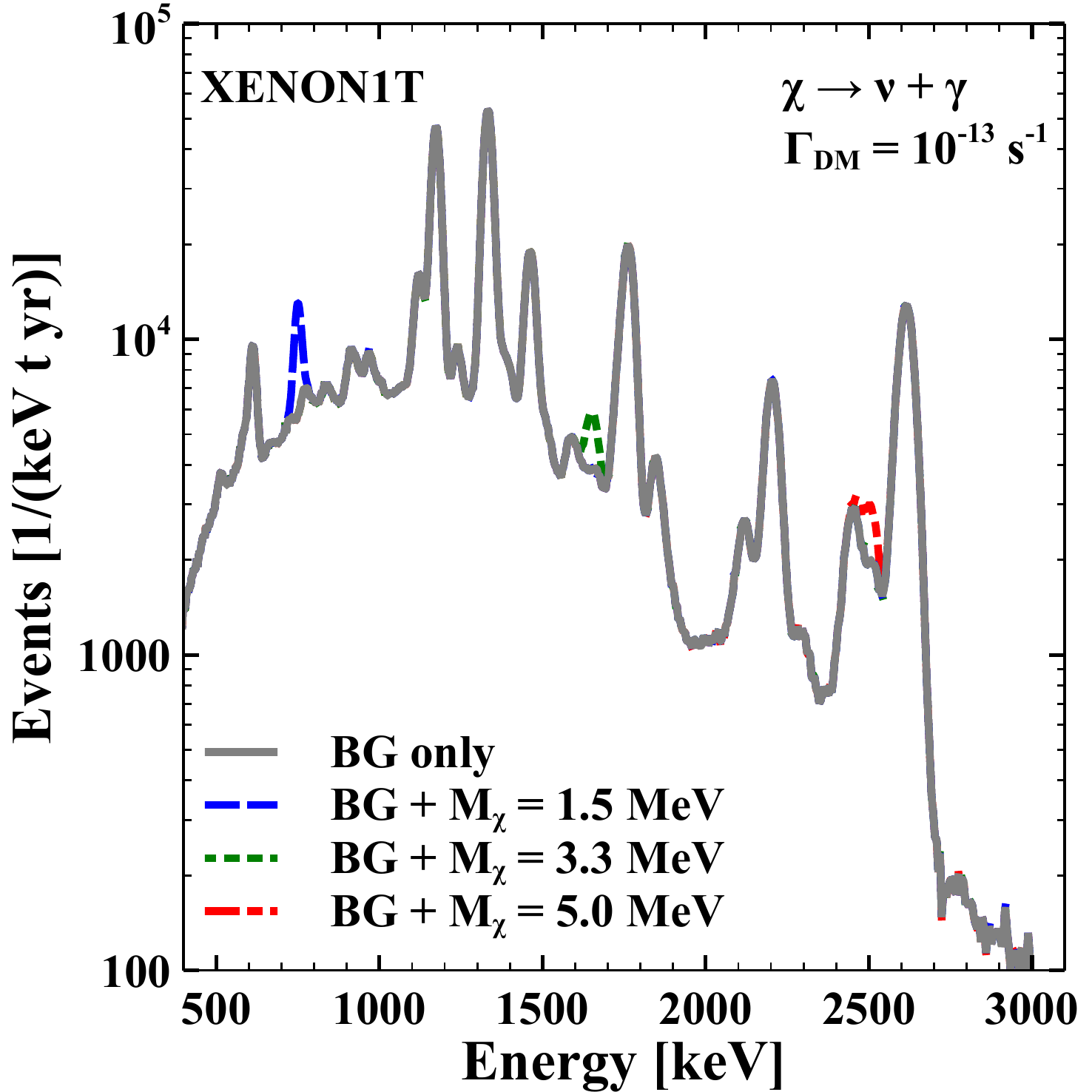} 
    \caption{The XENON1T background (from\,\cite{Aprile:2020yad,Aprile:2020tmw} as described in the text) and some typical signals. The left panel displays the background in log-scale between $1$ keV and $3$ MeV, while the right panel shows the same background   in linear scale and only between $400$ keV and $3$ MeV. Some representative signals from a DM decay via $\chi \to \nu + \gamma$ are shown, on top of the background, for different DM masses. 
    }
    \label{fig:signal_BG}
\end{figure}

\noindent
{\bf Background spectra:}\\
In all considered cases, the signal is a mono-energetic peak appearing on top of the experimental background.
We use the published background spectrum from XENON1T: from\,\cite{Aprile:2020tmw} at low energies and from\,\cite{Aprile:2020yad} above  $200$\,keV. The latter data (given in arbitrary units in the publication) is scaled to fit the low energy rate. 
We take the single scattering data of~\cite{Aprile:2020yad} below 350\,keV and the spectrum of multiple interactions above this energy since the probability of the $\gamma$-ray to undergo multiple scatters increases with energy.  
This is a conservative choice since the background for multiple scatterings is higher than the single scatter one above 350\,keV.
Figure\,\ref{fig:signal_BG} left shows, as an example, the signals from the decay $\chi \to \nu + \gamma$, for a 100, 200, and 400\,keV DM particle (blue, green, and red, respectively) on top of the background spectrum of XENON1T (in gray).  
Examples for the same decay channel are shown in figure\,\ref{fig:signal_BG} right for DM masses corresponding to 1.5\,MeV, 3.3\,MeV, and 5.0\,MeV.
The underlying DM decay rate for all examples is $\Gamma =10^{-13}\,\textrm{s}^{-1}$.
The sensitivity of the experiment depends on the specific DM mass due to the features of the background spectrum and the expected signal rates (see equations\,(\ref{eq:anni}, \ref{eq:decay})).
Several lines are visible in figure\,\ref{fig:signal_BG} originating from backgrounds in the detector, for example the 41\,keV line from $^{83m}$Kr or the 1.17\,MeV and 1.33\,MeV lines from $^{60}$Co decays. For masses corresponding to peaks in the background spectrum, the sensitivity can be up to an order of magnitude lower than for masses in-between. 
\vskip 0.2cm

To calculate the sensitivity of future DM detectors to DM annihilation and decay, we need to make some assumptions on their background levels. For XENONnT, we take the background shape from XENON1T and assume a background suppression factor of 8. We study however also the effect of a pessimistic (factor 5) and more optimistic (factor 10) background suppression. Similarly, we employ a factor 80 as suppression factor for the DARWIN background compared to the background of XENON1T and test the effect of factors 50 and 100.
\vskip 0.2cm

\begin{figure}[t]
    \centering
    \includegraphics[width=0.48\textwidth]{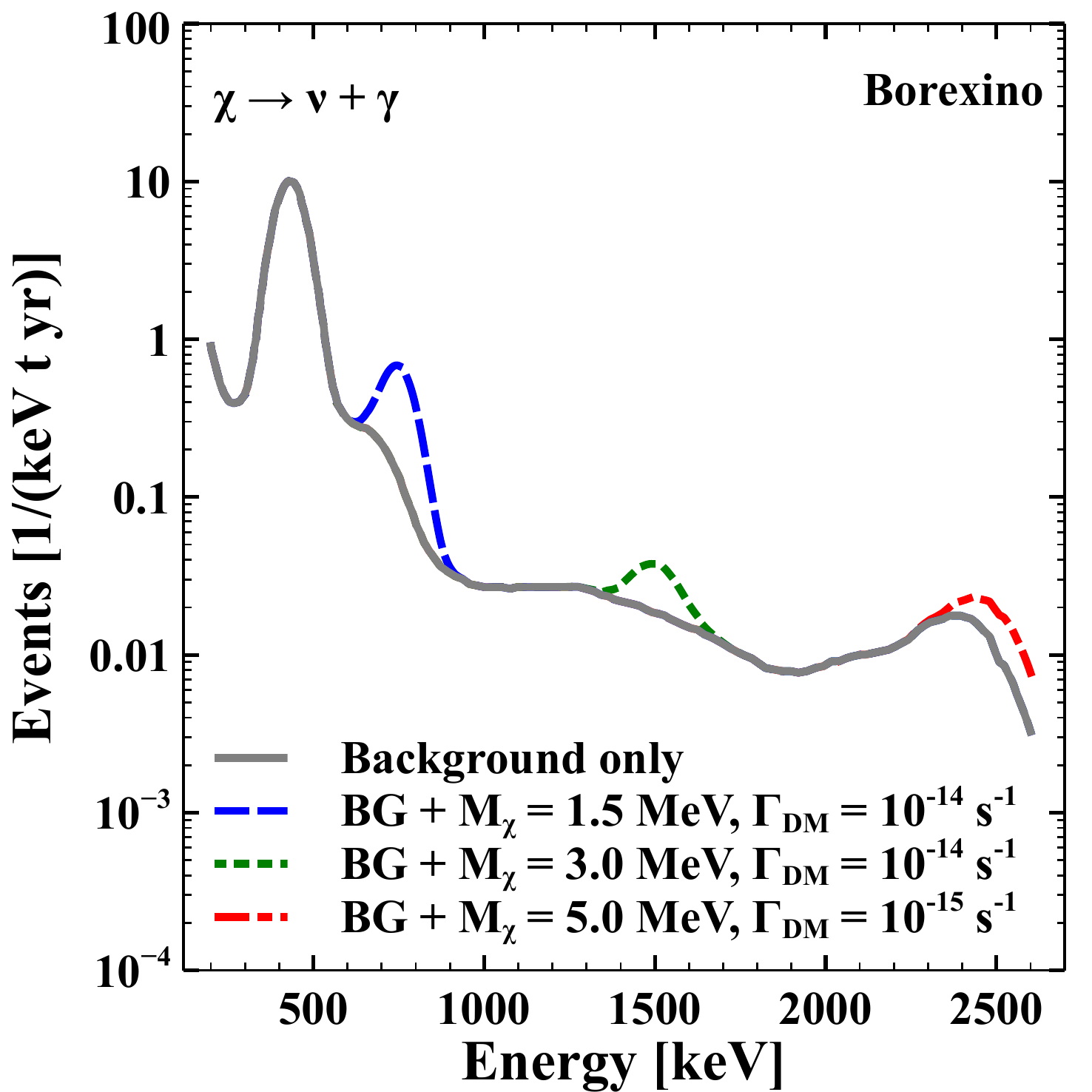} \hspace{3mm}
    \includegraphics[width=0.48\textwidth]{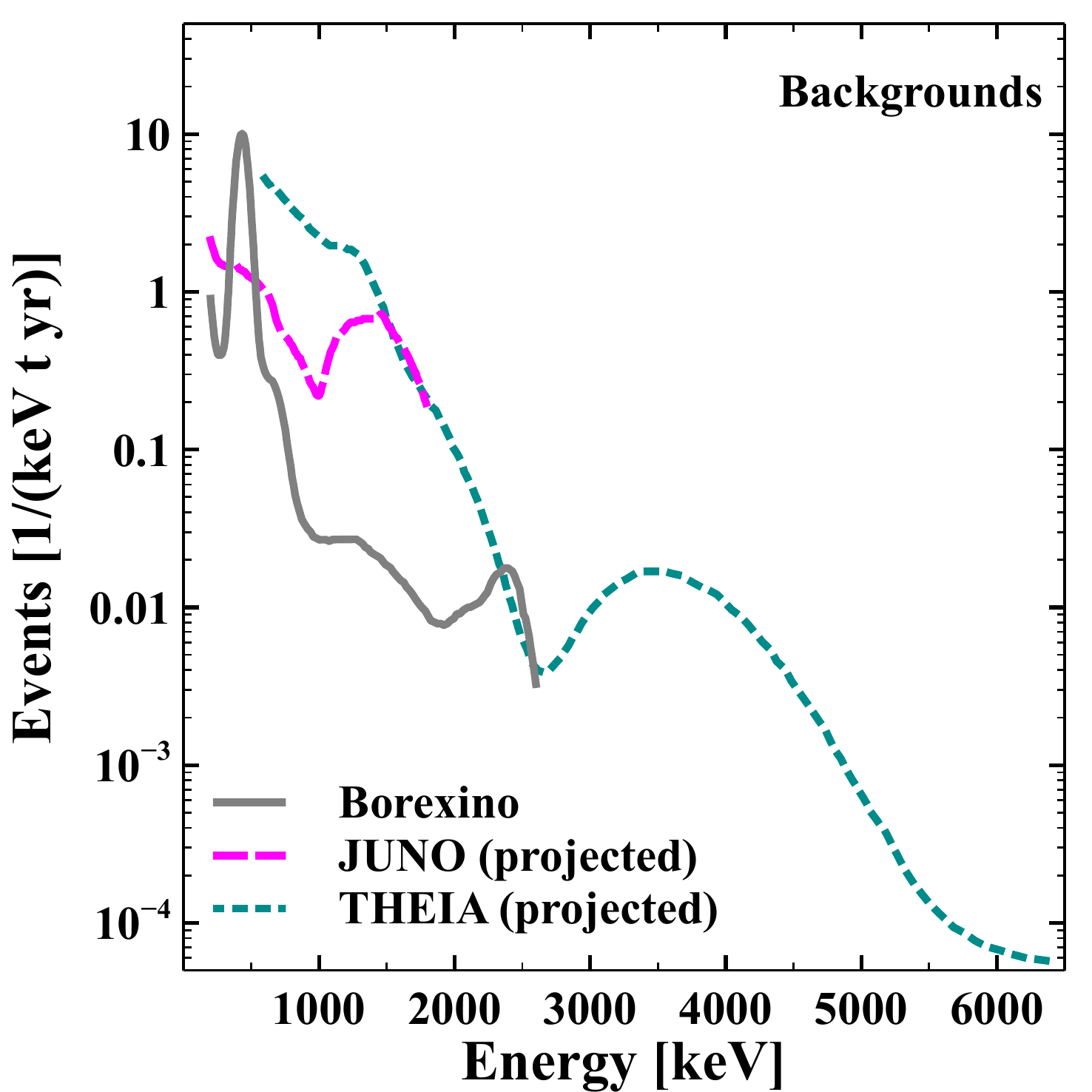} 
    \caption{Left: Borexino background spectrum (from\,\cite{Agostini:2018uly}) including some typical signals. The right panel shows the comparison of the Borexino spectrum to the expected spectra of JUNO and THEIA from\,\cite{Djurcic:2015vqa} and~\cite{Askins:2019oqj}, respectively.  
    }
    \label{fig:signal_BG_Borex}
\end{figure}

We take as well the published background spectrum of Borexino from~\cite{Agostini:2018uly} to derive results.
The left panel of figure\,\ref{fig:signal_BG_Borex} shows this spectrum together with some representative decay lines ($\chi \to \nu + \gamma$) for a DM mass of 1.5, 3.0, and 5.0\,MeV (in blue, green, and red respectively) and the decay rate is either $10^{-14}\,$s$^{-1}$ (blue, green) or $10^{-15}\,$s$^{-1}$ (red).
The structure of the spectrum is given by the sum of the various solar neutrino components ($^7$Be and $^8$B for instance) and some backgrounds from natural radioactivity in the detector materials.
The sensitivity for a signal depositing around $\sim 500$\,keV is significantly affected by the $^{210}$Po $\alpha$-decay peak at this energy.
The energy threshold of Borexino is significantly higher than for the xenon-based experiments because the background from $^{14}$C present in the organic liquid becomes dominant below $\sim 150$\,keV (see the rise of the rate in figure\,\ref{fig:signal_BG_Borex}). 
\vskip 0.2cm

For JUNO, we use the published expected spectrum (top panel of figure 1.10 in\,\cite{Djurcic:2015vqa}) which starts close to the end-point of the $^{14}$C spectrum and covers up to 1.8\,MeV where the background from $^{11}$C dominates. In the case of THEIA, we take the expected background spectrum for a 25\,ktonne target from figure 8 in\,\cite{Askins:2019oqj}. This spectrum starts at a bit higher energy and ranges up to 6\,MeV. The peak at $\sim 4$\,MeV originates from $^{208}$Tl of the thorium chain inside the scintillator including the 2.6\,MeV $\gamma$-line and the energy deposition from the associated beta decay. Below this energy, signals from $^7$Be neutrinos and $^{40}$K background decays determine the shape of the spectrum.  The used spectra for JUNO and THEIA are shown in the right panel of figure\,\ref{fig:signal_BG_Borex}.
\vskip 0.2cm

The precise knowledge of the background is one of the main advantages of the laboratory searches we are proposing here. In astrophysical environments, it is essentially impossible to measure the background. Thus, given a putative signal, it is extremely difficult to ascertain that it is actually due to DM. The long-standing controversies regarding the Galactic Center excess (see e.g.\ \cite{Goodenough:2009gk,Bartels:2015aea}) and the 3.5\,keV line (see e.g.\ \cite{Boyarsky:2014ska,Jeltema:2014qfa}), for instance, are mostly due to an uncertain background.
\vskip 0.2cm

\section{\label{sec:proc_results}Results}
\noindent

In this section, we show the obtained limits for the annihilation cross section or the decay rate for each of the final states considered, and the projected sensitivity of future experiments.
A $\chi^2$-method is employed to test the signal hypothesis.
For this purpose, we take the background spectrum as described in section\,\ref{sec:Response and BG} and test for each DM mass the allowed signal-strength (annihilation cross section or decay rate) of the signal models discussed in section\,\ref{sec:signal}.
In all cases, the signal is modeled as a Gaussian peak with a width given by the corresponding energy resolution.
Details regarding the features of the DM annihilation and decay signals can be found in section\,\ref{sec:res_annihilation} and section\,\ref{sec:res_decay}.
We employ a $\chi^2$ probability density function (pdf) with one degree of freedom, 
\begin{equation}
    p\left(x\right)=\frac{1}{\sqrt{2\pi}}~ x^{-1/2} \cdot e^{-x / 2},
    \label{eqn:chi2pdf}
\end{equation}
to approximate the log-likelihood ratio test-statistics of a binned Poisson hypothesis test with one free parameter in the signal model \cite{Cowan:2010js}.
\vskip 0.2cm

As a threshold for the upper limit, we take the 90\% quantile of the $\chi^2$ pdf in equation\,(\ref{eqn:chi2pdf}).
This threshold is independent of the tested signal strength (Wilk's theorem) which was verified for the obtained limits.
The threshold is compared to the computed $\chi^2$ of the signal-hypothesis under consideration, 
\begin{equation}
    \chi^2 =\sum_{i=0}^{N_{\text{bins}}} \frac{(O_i - E_i)^2}{E_i},
    \label{eqn:chi2}
\end{equation}
where $O_i$ corresponds to the bin-content when a signal is present and $E_i$ corresponds to the bin-content without signal with a 1\,keV binning.
This approximates a log-likelihood ratio scan of a binned Poisson hypothesis test and takes into account the statistical uncertainty in each bin.
The intersection between the 90\% quantile of the $\chi^2$ pdf and the scan through $\chi^2$-values per mass point as computed in equation\,(\ref{eqn:chi2}) denotes the upper limit that we set.

\vskip 0.2cm

\subsection{\label{sec:res_annihilation}Dark Matter Annihilation}

\begin{figure}[h]
    \centering
    \includegraphics[width=0.32\textwidth]{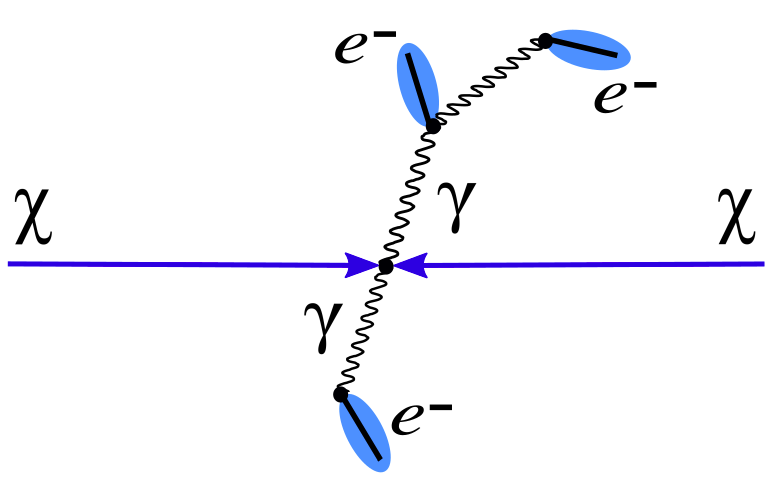}
    \caption{Schematic of the annihilation reaction $\chi +\chi  \to \gamma + \gamma$, where one of the $\gamma$'s (bottom) undergoes photo-absorption and the second (top) experiences a Compton scattering before being absorbed. The electrons created deposit their kinetic energy leading to the detected signals.}
    \label{fig:SchemeGG}
\end{figure}

$\boldsymbol{\chi + \chi  \to \gamma + \gamma}$:
The $\gamma$-rays produced in this annihilation reaction can only be observed indirectly once they interact with electrons in the medium. We therefore observe the electrons produced via the photo-effect or Compton effect. Figure \ref{fig:SchemeGG} shows schematically a possible event topology.

The mass of the DM particles corresponds to the energy of the $\gamma$-rays and therefore the total energy deposited is:
\begin{equation}
E_{{\rm dep}}^{\chi \chi  \to \gamma \gamma} = 2\cdot m_{\chi}.
\end{equation}
For low energies, an electron is emitted mostly in the photo-effect process and deposits its energy after traveling a short distance. With increasing energy, the probability to undergo multiple scatterings grow. Indeed, the XENON1T data shows (figure\,6 in~\cite{Aprile:2020yad}) that the background rate of multiple scatterings is higher for energies above 350\,keV. Consequently for liquid xenon detectors, the single scattering spectrum is employed below 350\,keV and the multiple scattering spectrum above that energy, as explained in Sec.\,\ref{sec:Response and BG}.
\vskip 0.2cm

\begin{figure}[t]
    \centering
\includegraphics[width=0.47\textwidth]{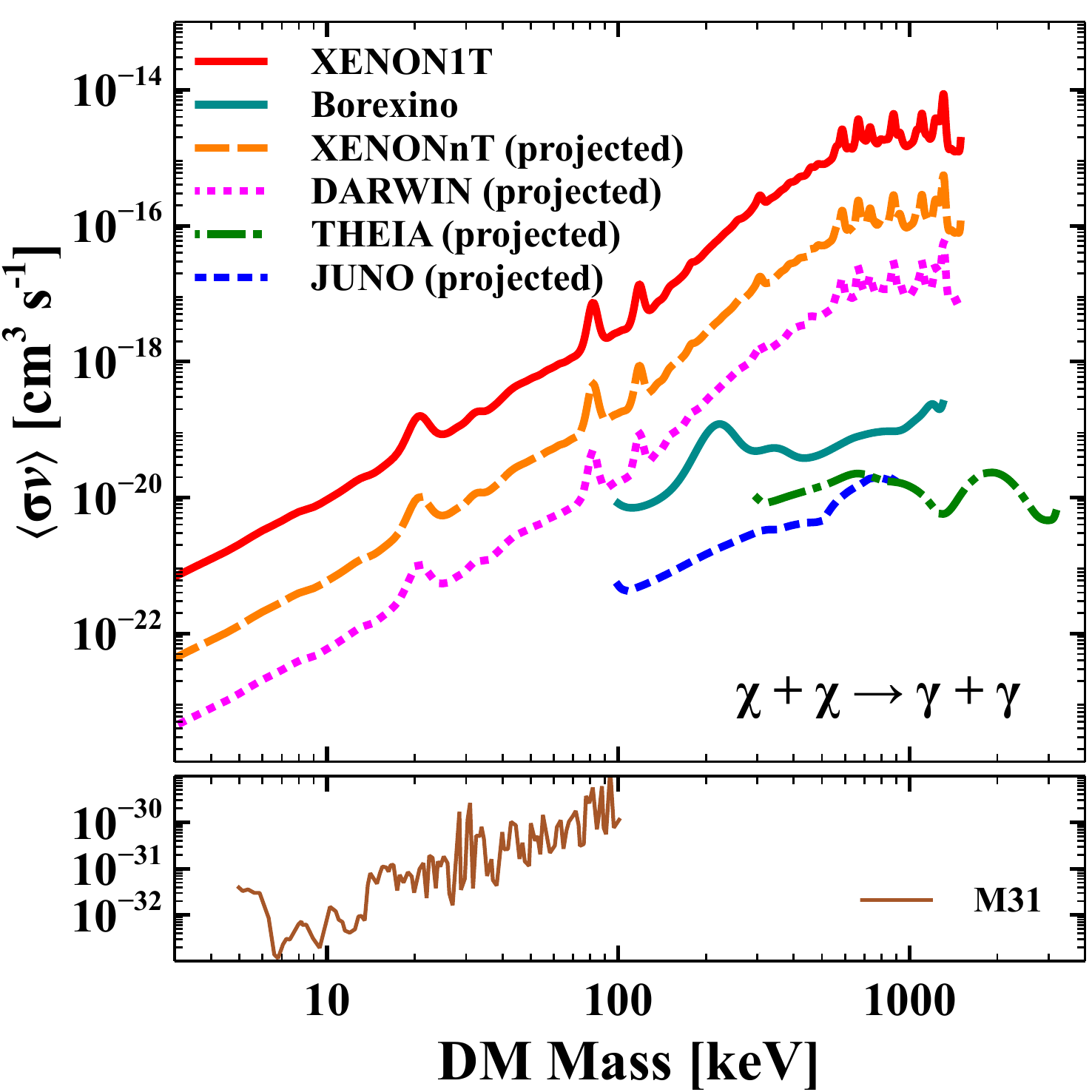}
\includegraphics[width=0.47\textwidth]{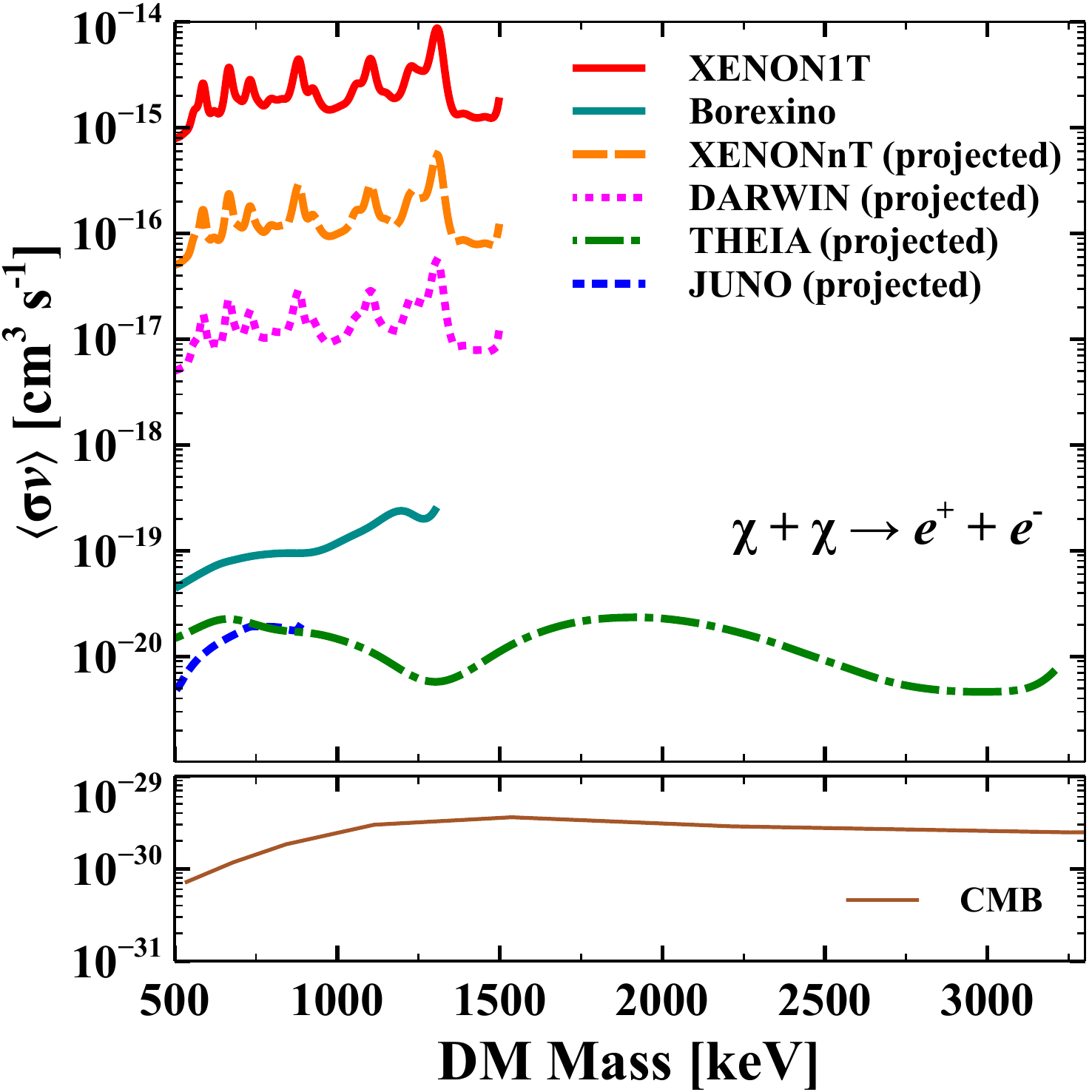}
    \caption{Limits and prospects  for $\chi + \chi \rightarrow \gamma +\gamma$ (left) and $\chi + \chi \rightarrow e^{-} + e^{+}$ (right).
    The red line (solid) shows the derived limit from XENON1T whereas the orange (dashed) and magenta (dotted) lines correspond to the expected sensitivities of XENONnT and DARWIN respectively.
    The computed limits for Borexino, THEIA, and JUNO are shown in aquamarine (solid), green (dashed-dotted), and blue (dashed) respectively.
    Notice that in the right panel the range of DM masses  is smaller (because $m_{\chi}>m_e$) and the scale is linear rather than logarithmic. For comparison, the lower panels show the corresponding bounds from M31 \cite{Ng:2019gch} or the CMB \cite{Slatyer:2015jla}.
    }
    \label{fig:res_anni_gammagamma}
\end{figure}

The results for this case are shown in the left panel of figure\,\ref{fig:res_anni_gammagamma}. The region above the solid red line is excluded by the data from XENON1T.
The dashed (orange) and dotted (magenta) lines show instead the expected sensitivities of XENONnT and DARWIN which are predicted to strengthen XENON1T limit by factors of about $20$ and $200$ respectively over the entire mass range.
As expected from equation\,(\ref{eq:anni}), the limits decrease quadratically with the DM mass.
On top of this general trend, the effect of the bumps present in the background is clearly visible.
Current limits of XENON1T on $\langle \sigma v\rangle$, then, lie between $10^{-21}\,\mathrm{cm^3\,s^{-1}}$ at a mass of  $4$\,keV and $10^{-15}\,\mathrm{cm^3\,s^{-1}}$ at a $2$\,MeV mass. 
\vskip 0.2cm

The limits derived from Borexino data, significantly more stringent, are shown by a continuous aquamarine line in figure\,\ref{fig:res_anni_gammagamma}. 
Borexino constrains $\langle \sigma v\rangle$ to be below  between $10^{-19}\,\mathrm{cm^3\,s^{-1}}$ at a DM mass of 1\,MeV. 
The expected limits of JUNO and THEIA are shown in  blue and green dashed-lines, respectively, and are expected to be one to two orders of magnitude stronger mainly due to their higher exposures.
The shape of the limit curves for these experiments can also be understood by the shape of the background spectra in figure\,\ref{fig:signal_BG_Borex} and the dependence on $1/m_\chi^2$ of the annihilation rate.
The larger volume of Borexino, THEIA and JUNO (see table\,\ref{tab:summ}) implies that the limits are stronger, but note that the xenon-based experiments can probe much smaller DM masses, because of their much smaller energy threshold.
The liquid xenon experiments need a $\sim$\,keV energy threshold in order to directly detect DM.
Above $\sim 100$\,keV and depending on the DM mass, the constraints from Borexino can be $(3-5)$ orders of magnitude better than XENON1T. 
\vskip 0.4cm

\begin{figure}[h]
    \centering
    \includegraphics[width=0.32\textwidth]{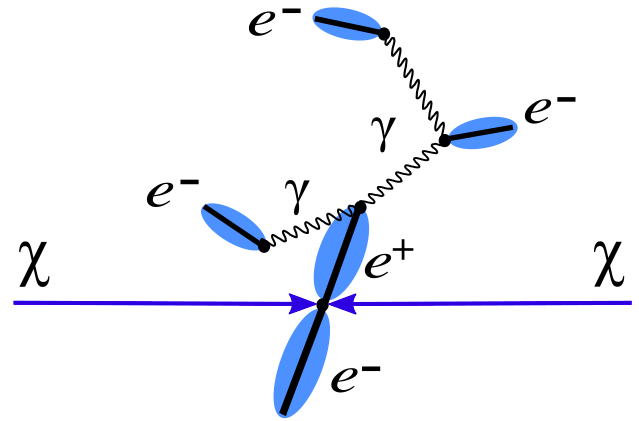}
    \caption{Schematic of the annihilation reaction $\chi +\chi  \to e^- + e^+$. The positron deposits its energy and annihilates with an electron of the medium producing two 511\,keV $\gamma$-rays. These interact promptly with the medium depositing also their energy.}
    \label{fig:SchemeEE}
\end{figure}

$\boldsymbol{\chi + \chi  \to e^- + e^+}$:
For the annihilation into electron and positron, the two charged leptons deposit their energy in the medium. In addition, once the $e^+$ is slowed down, it annihilates with an electron emitting two 511\,keV $\gamma$-rays. Figure \ref{fig:SchemeEE} shows a scheme of the annihilation reaction.

In this case, the minimum DM mass that can be tested is 511\,keV as this is the mass required to create an electron-positron pair. The total energy deposited corresponds also to
\begin{equation}
E_{{\rm dep}}^{\chi \chi  \to e^- e^+} = 2\cdot m_{\chi},
\end{equation}
 because the energy required for the creation of the two leptons is obtained back when the positron annihilates with an electron at rest. For the liquid xenon case, we employed a simplified simulation to study the energy resolution resulting from the depositions of the $e^-  e^+$ and the $\gamma$'s. We found that the resolution can be modeled by taking the multiple scattering curve for the total energy as described in Sec.\,\ref{sec:Response and BG}. With this choice, we obtain conservative results as the actual resolution is slightly better.

\vskip 0.2cm

The right panel of figure\,\ref{fig:res_anni_gammagamma} displays our results for this case.  Notice that the mass range is much smaller and that the scale is linear rather than logarithmic. Since there is a kinematic threshold for the DM mass in this channel, Borexino, JUNO and THEIA give better limits than the xenon-based experiments for all masses. The fact that the limits turn out to be essentially flat over the entire mass range is a consequence of the decreasing background -- see the right panel of figures\,\ref{fig:signal_BG} and \ref{fig:signal_BG_Borex}. The limits and projections obtained are seen to be comparable to those from the $\gamma\gamma$ final state at the same masses.

\subsection{\label{sec:res_decay}Dark Matter Decay}
We have seen in the previous subsection how the shape of the limit curves can be understood from the shape of the background spectra, the  volume of the experiments and the $1/m_\chi^2$ 
mass-dependence  of the rate. For decay, everything can be interpreted in analogy, with the exception that the rate scales with $1/m_\chi$. 
For DM decay, there is an unavoidable limit in terms of the lifetime $t_0$ of the Universe. Some of our limits will lie above this value and are hence unphysical. This may be expected giving the comparably small volume of the terrestrial experiments. Our approach is to use existing and expected data to set limits, so we nevertheless will display the results above this unavoidable value. 

\vskip 0.2cm

\begin{figure}[th!]
    \centering
\includegraphics[width=0.47\textwidth]{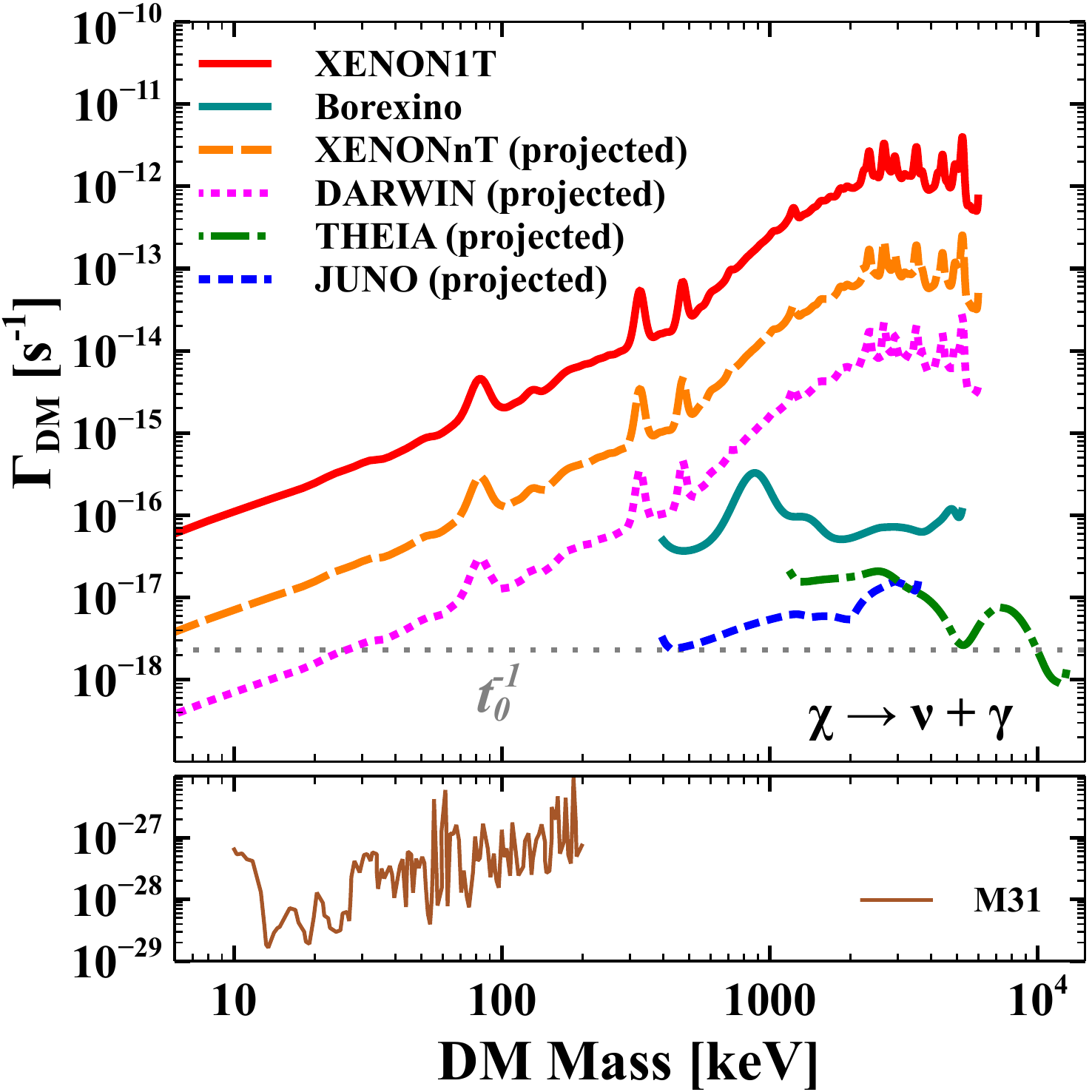}
\includegraphics[width=0.47\textwidth]{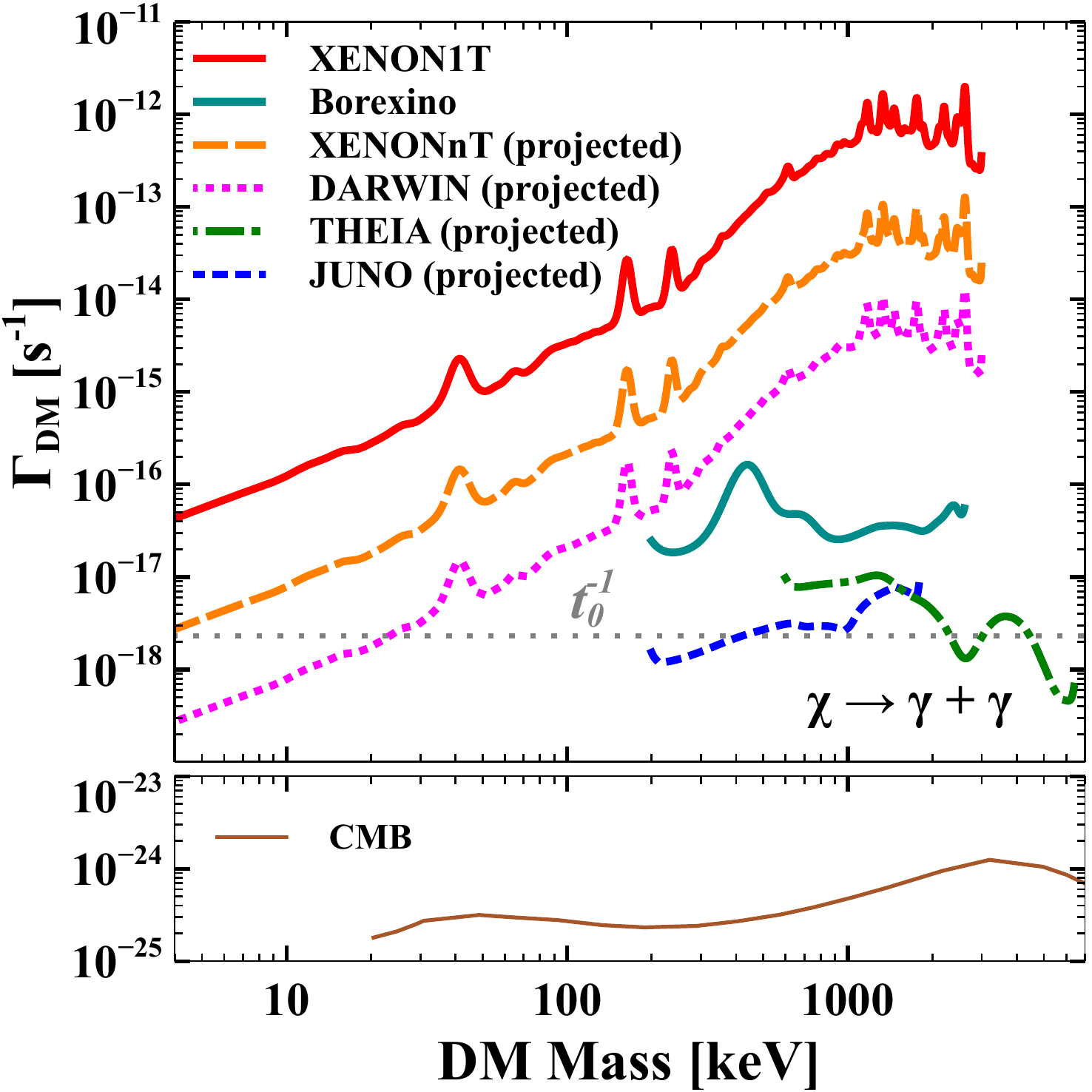}\\[5mm]
\includegraphics[width=0.47\textwidth]{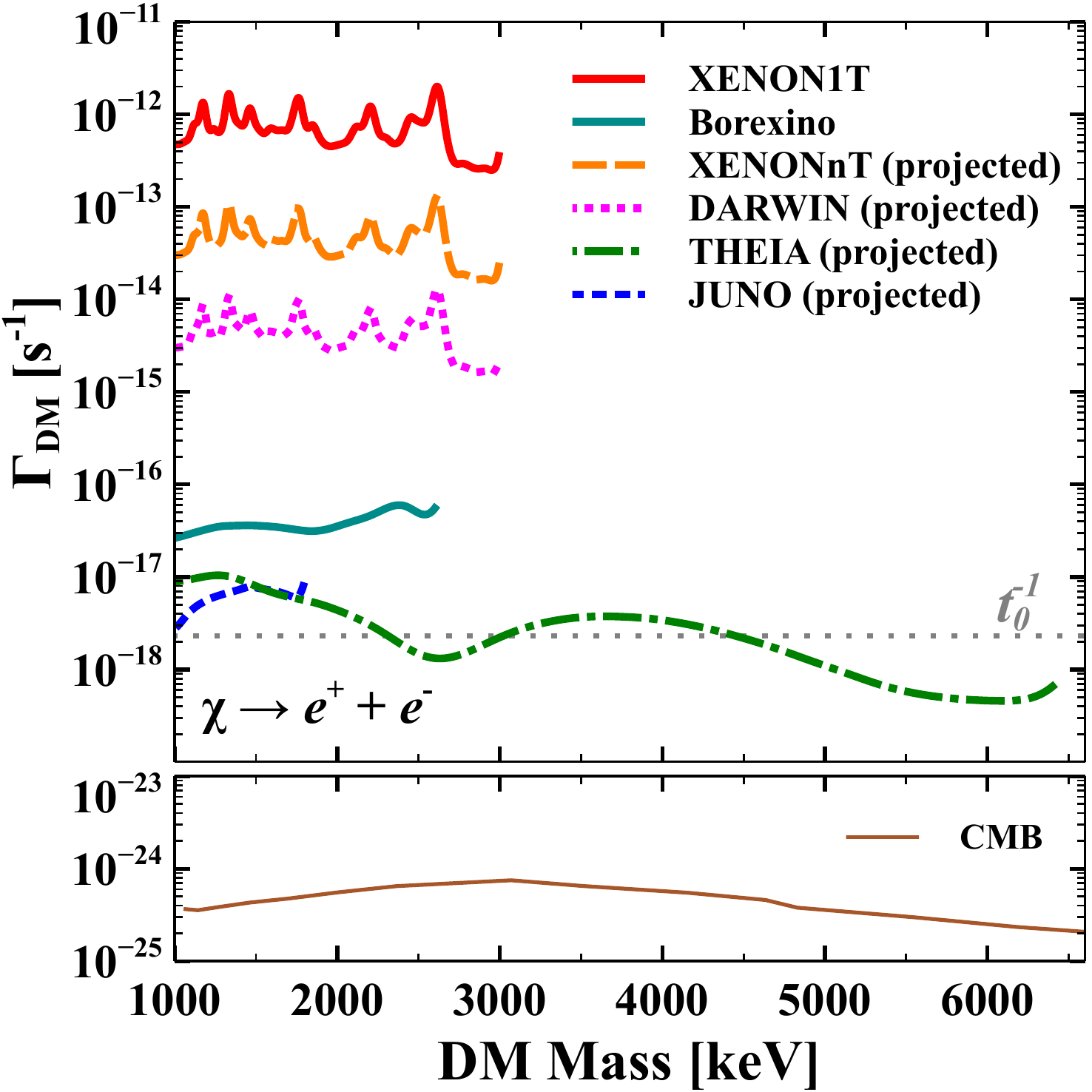}
    \caption{Limits and prospects  for $\chi \rightarrow \nu + \gamma$ (top left), $\chi  \to \gamma + \gamma$ (top right) and $\chi  \to e^- + e^+$ (bottom). The solid lines show the results from XENON1T (red) and Borexino (aquamarine). The projections for XENONnT and DARWIN are shown in dashed (orange) and dotted (magenta) lines respectively, while those for THEIA and  JUNO are displayed as dashed-dotted (green) and dashed (blue) lines. Notice that in the bottom panel the range of DM masses  is smaller (because $m_{\chi}>2m_e$) and the scale is linear rather than logarithmic. The horizontal dotted (gray) line corresponds to a lifetime equal to the age of the Universe, $t_0$. For comparison, the lower panels show the respective bounds from M31 \cite{Ng:2019gch} or the CMB \cite{Slatyer:2016qyl}}.
    \label{fig:res_decay_nugamma}
\end{figure}

$\boldsymbol{\chi  \to \nu + \gamma}$:
This is the most important decay mode for sterile neutrinos, the archetype DM candidate at keV masses.  In this decay mode, the neutrino escapes the detector and only the $\gamma$-ray is detected. The energy deposited by the $\gamma$-ray corresponds in this case to half of the DM mass:
\begin{equation}
E_{{\rm dep}}^{\chi  \to \nu \gamma} = m_{\chi}/2.
\end{equation}
Our results for this case are shown in the top-left panel of figure\,\ref{fig:res_decay_nugamma}. The limits on $\Gamma_{\rm DM}$ tend to weaken linearly with the DM mass, according to equation\,(\ref{eq:decay}) and up to background effects.  The actual limit on $\Gamma_{\rm DM}$ from XENON1T varies from about $10^{-17}\,$s$^{-1}$ at a mass of few keV to $10^{-13}\,$s$^{-1}$ at a mass of few MeV. 
In DARWIN, these limits may be improved by about three orders of magnitude.
Also here, as the xenon-based experiments have a lower energy threshold, they can test much lower DM masses while Borexino, JUNO and THEIA provide in turn better limits for the higher energy range. For instance, Borexino gives a limit of $10^{-19}\,$s$^{-1}$ at a mass of 1\,MeV, which can be improved by an order of magnitude by THEIA and JUNO. 

\vskip 0.4cm

$\boldsymbol{\chi  \to \gamma + \gamma}$:
The mass of the DM particle is transferred to kinetic energy of the $\gamma$-rays. Therefore, the total energy deposition of the $\gamma$'s corresponds to the DM mass:

\begin{equation}
E_{{\rm dep}}^{\chi  \to \gamma \gamma} = m_{\chi}.
\end{equation}
The top-right panel of figure\,\ref{fig:res_decay_nugamma} shows the current limit and future prospects for this decay.
The results compare to each other similarly to the results of the $\gamma +  \nu$ decay channel.
\vskip 0.4cm

$\boldsymbol{\chi  \to e^- + e^+}$:
This decay mode is similar to the annihilation into $e^-  e^+$. The two leptons deposit their kinetic energy and once stopped, the positron annihilates with an electron producing two 511\,keV $\gamma$-rays. The total energy deposition corresponds in this case to the DM mass: 
\begin{equation}
    E_{{\rm dep}}^{\chi  \to e^-  e^+} = m_{\chi}.
\end{equation}
While the limit of XENON1T on $\Gamma_{\rm DM}$ lies at the level of  $10^{-13}\,$s$^{-1}$, for Borexino the limit is significantly lower at the level of a few times $10^{-17}\,$s$^{-1}$. The future JUNO and THEIA detectors would improve Borexino results by an order of magnitude.
\vskip 0.2cm

For all annihilation and decay channels described above, the effect of the background on the results of the upcoming xenon experiments has been studied.
We assume background suppression factors of 8 and 80 for XENONnT and DARWIN, respectively.
 This is also the background level used for all results of the figures above.
 We test the effect of different background level into the results and find a small changes in the obtained limits.
 For a pessimistic scenario (factors 5 and 50), the limits worsen by 12\%, while for an optimistic scenario (factors 10 and 100), the limits  improve by 25\% for both XENONnT and DARWIN.

\section{\label{sec:dis}Discussion and Conclusion}
\noindent
Large dark-matter and neutrino detectors have a rich experimental programs beyond their main scientific goals. We have reported here on a new  possibility in which the decay or annihilation of DM particles may take place inside the Earth-based detectors. Focusing on XENON1T and Borexino as well as the future XENONnT, JUNO, DARWIN and THEIA, we have set current and prospective future limits on annihilation/decay rates for DM masses in the keV to few MeV range. The relevant final states thus involve  electrons, positrons and photons. Usually, cosmological considerations \cite{Slatyer:2015jla} or observations of astrophysical objects \cite{Ng:2019gch} are used to set such limits, and they  turn out to be more stringent, by several orders of magnitude, than the ones we derived. Nevertheless, the limits obtained here under well-controlled laboratory conditions offer the benefits of a well-known background composition. Moreover, the signal  does not suffer from uncertainties associated with the DM distribution -- the $J$- and $D$-factors. While the astrophysical and cosmological limits are not expected to improve significantly, upcoming direct detection and neutrino experiments will improve the laboratory limits and close the gap considerably. Indeed at low energies (below $\sim100$\,keV),  the upcoming XENONnT and DARWIN experiments will improve the limit of XENON1T by at least an order a magnitude each. Similarly above this energy, the larger detectors JUNO and THEIA will improve the results of Borexino by an order of magnitude.
\vskip 0.4cm

Similar studies could be performed in argon dark-matter detectors like DarkSide and Argo\,\cite{Aalseth:2017fik}. 
Moreover, Super-Kamiokande\,\cite{Fukuda:2002uc} and Hyper-Kamiokande\,\cite{Abe:2018uyc} have much larger volumes and could be employed for similar studies at MeV-GeV energies. DUNE\,\cite{Abi:2020loh} or large-scale experiments looking for neutrinoless double beta decay such as KamLAND-Zen\,\cite{KamLAND-Zen:2016pfg} are other up-coming facilities with features that allow to set limits along the lines proposed here. 
The huge detectors at LHC, which are optimized for energies in the TeV-range, could also provide limits for some final states and DM masses. All such studies, however, lie beyond the scope of present paper. 
Our focus was here on final states involving photons, electrons and positrons. In principle, for larger DM masses also other particles could be produced, for instance $\mu^-$ and $\mu^+$. While being a spectacular signature, energy deposition by the muons followed by a delayed electron/positron signal from the two muon decays), the background and efficiencies need to be carefully evaluated before analyzing this further. The same holds for other final states. Of course, with increasing DM mass the amount of particles in the detector volume decreases. 
\vskip 0.2cm

In summary, we have suggested a new method to test DM candidates using direct-detection and neutrino experiments as indirect detection laboratory experiments. We obtained present and future constraints on the decay and annihilation rates of DM particles.

\begin{acknowledgments}
\noindent We thank our colleague Michael Wurm for his valuable help. CY would like to thank Kenneth Garcia for his coding help in the initial stages of this work.
\end{acknowledgments}

\bibliography{bib}
\end{document}